\documentclass[final,1p,times]{elsarticle} 
\usepackage{graphicx} 
\usepackage{amssymb} 
\usepackage{amsthm} 
\usepackage{lineno} 

\journal{Nuclear Physics A} 
\begin{document} 

\begin{frontmatter} 


\title{Signals of the QCD Critical Point in Hydrodynamic Evolutions}

\author{Chiho Nonaka$^a$, M. Asakawa$^b$, S. A. Bass$^c$,  B. M\"uller$^c$}

\address{
$^a$Department of Physics, Nagoya University, Nagoya 464-8602, Japan \\
$^b$Department of Physics, Osaka University, Toyonaka 560-0043, Japan\\
$^c$Department of Physics, Duke University, Durham, North Carolina 27708, USA
}

\begin{abstract} 
The presence of a critical point in the QCD phase diagram can deform the trajectories describing the 
evolution of the expanding fireball in the $\mu_B$-$T$ phase diagram. 
The deformation of the hydrodynamic trajectories will change the transverse velocity ($\beta_T$) dependence of 
the proton-antiproton ratio when the fireball passes in the vicinity of the critical point. 
An  unusual $\beta_T$ dependence of the $\bar{p}/{p}$ ratio in a narrow beam energy window would
 thus signal the presence of the critical point.  
\end{abstract} 

\end{frontmatter} 



\section{Towards quantitative analyses of QCP}
The existence and  location of the QCD critical point (QCP) in the QCD phase diagram have been 
attracting  many physists'  interests in heavy ion collision physics. 
However recent studies based on effective theories show many possible location of the QCP in the 
QCD phase diagram. 
In addition, the latest finite temperature and imaginary chemical potential lattice QCD calculation 
shows that even the existence of the QCD critical point is uncertain \cite{QCDpd}. 
At present experiments and quantitative  phenomenological 
analyses for the QCP are needed, because it seems to be very difficult 
to reach  a solid conclusion about the QCP just from lattice QCD and effective theories.  
Here we discuss the consequences of the QCP from point of view of the quantitative phenomenological analyses in 
heavy ion collisions. 
Towards quantitative analyses of the QCP we need 
the following  three steps: a realistic dynamical model which describes expansion of  hot and dense matter 
after collisions, an equation of state (EoS) with 
QCP  and appropriate physical observables which show signals of QCP clearly.  

For a realistic dynamical model,  we use a combined
fully three-dimensional macroscopic / microscopic transport approach
employing relativistic 3d-hydrodynamics for the early, dense, deconfined stage of the reaction and a
microscopic non-equilibrium model for the later hadronic stage where the equilibrium assumptions
are not valid anymore \cite{NoBa07}. Within this approach we study the dynamics of hot, bulk QCD matter,
which is being created in ultra-relativistic heavy ion collisions at RHIC. Our approach is capable of
self-consistently calculating the freezeout of the hadronic system, while accounting for the collective
flow on the hadronization hypersurface generated by the QGP expansion. 
We succeed in explaining a lot of experimental data consistently at RHIC: $P_T$ spectra of various particles including multistrangeness 
particles,  rapidity distributions, mean $P_T$ as a function of particle mass, freezeout time distribution of particles, 
and elliptic flow. 
In this calculation, we adopt  the EoS with the 1st order phase transition without QCP which is used in most hydrodynamic models. 
Now we replace the EoS in 3d hydro + UrQMD model by an EoS with QCP. 

 The EoS of QCP which we construct here is composed of two parts: one is a singular part 
around QCP and another part is non-singular part which is described by usual 
QGP phase and hadron phase \cite{NoAs05}. 
For the singular part  of the EoS, we assume that QCD has the same universality class 
as 3d Ising model. 
First we construct the EoS of 3d Ising model as a function of reduced temperature ($r$) 
and external magnetic field ($h$) \cite{GuZi97} and map it on the QCD phase diagram, $\mu_B$-$T$ plane.  
The magnetization as a functions of $r$ and $h$ in 3d Ising model shows  different behavior of phase transition between 
negative $r$ (1st order) and positive $r$ (crossover).
In mapping  the EoS with 3d Ising model as a function of $r$ and $h$ on the $\mu_B$-$T$ plane, 
however, there is no universality between QCD and 3d Ising model.  
We can not determine the direction of $h$ axis to $r$ axis, though we can fix $r$ axis as it  is parallel to tangential line 
at QCP  to the phase boundary.  Here we choose the direction of $h$ to be perpendicular to $r$ axis. 
Besides not only the size of critical region around the QCD and location of it on the QCD phase diagram 
are input parameters in our model.   
The EoS with the QCP has a very interesting feature in isentropic trajectories  on $\mu_B$-$T$ plane: 
the QCP works as an attractor of isentropic trajectories on the QCD phase diagram \cite{StRaSh98, NoAs05}, which 
gives us a clue of finding the QCP. 

\section{Signals of QCP}
Next we discuss how to find clear consequences of QCP in heavy ion collisions. 
The promising signals of QCP should survive not only in expanding fireball but also 
even after the freezeout process. 
Here we investigate two candidates of them: fluctuations and hadron ratios.  
For fluctuation, naively, we have to pick up fluctuations of conserved values such as  charge, baryon number 
and strangeness during whole process of collisions.  
Hadron ratios are fixed at chemical freezeout temperature and hold the same value during 
freezeout process and final state interactions.  

\begin{figure}[h]
\begin{minipage}{0.45\textwidth}
\includegraphics[width=\textwidth]{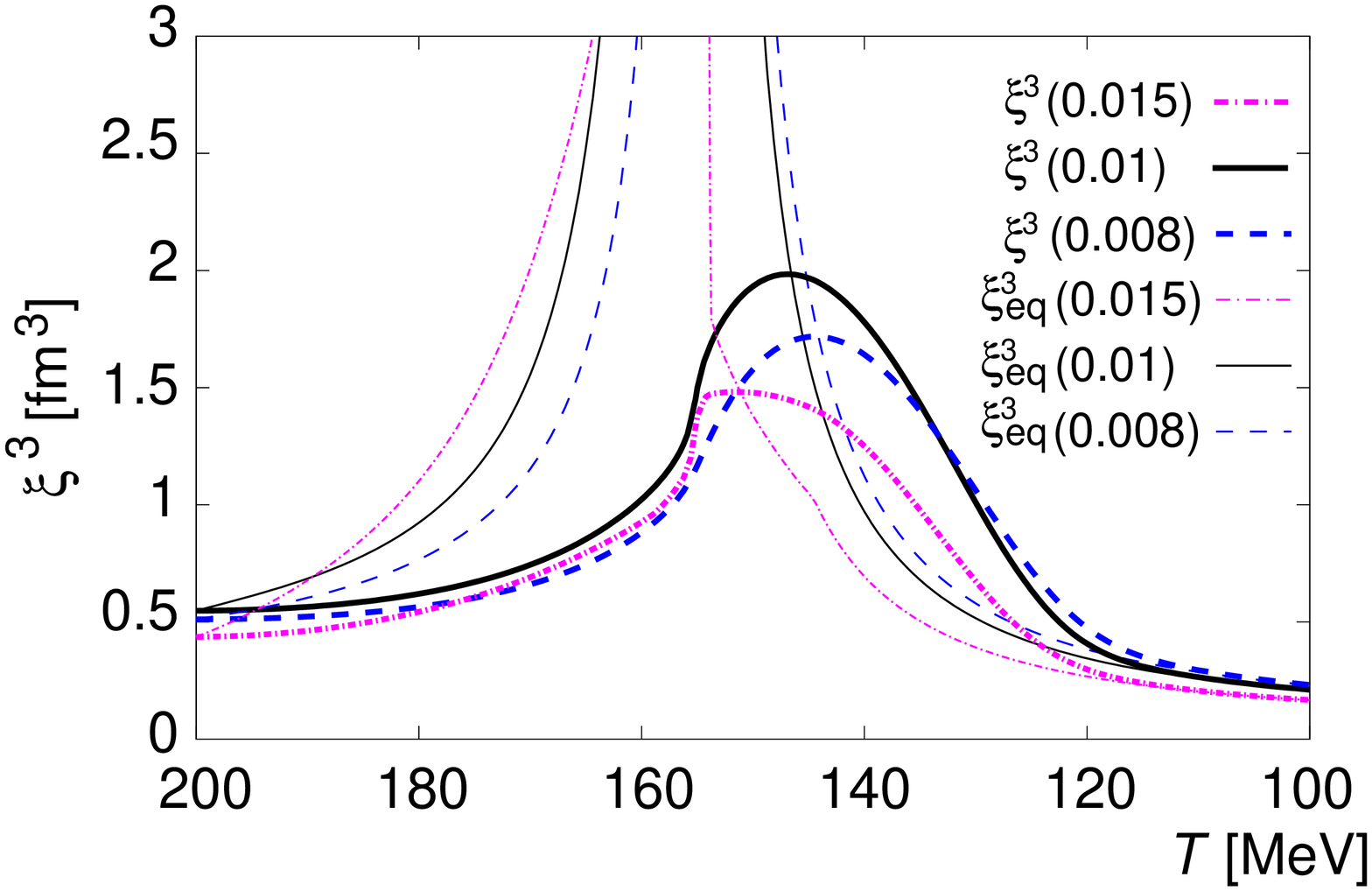}
\caption[]{Fluctuations as a function of temperature along isentropic trajectories. 
}
\label{Fig-xi}
\end{minipage}
\hspace{0.04\textwidth}
\begin{minipage}{0.45\textwidth}
\includegraphics[width=\textwidth]{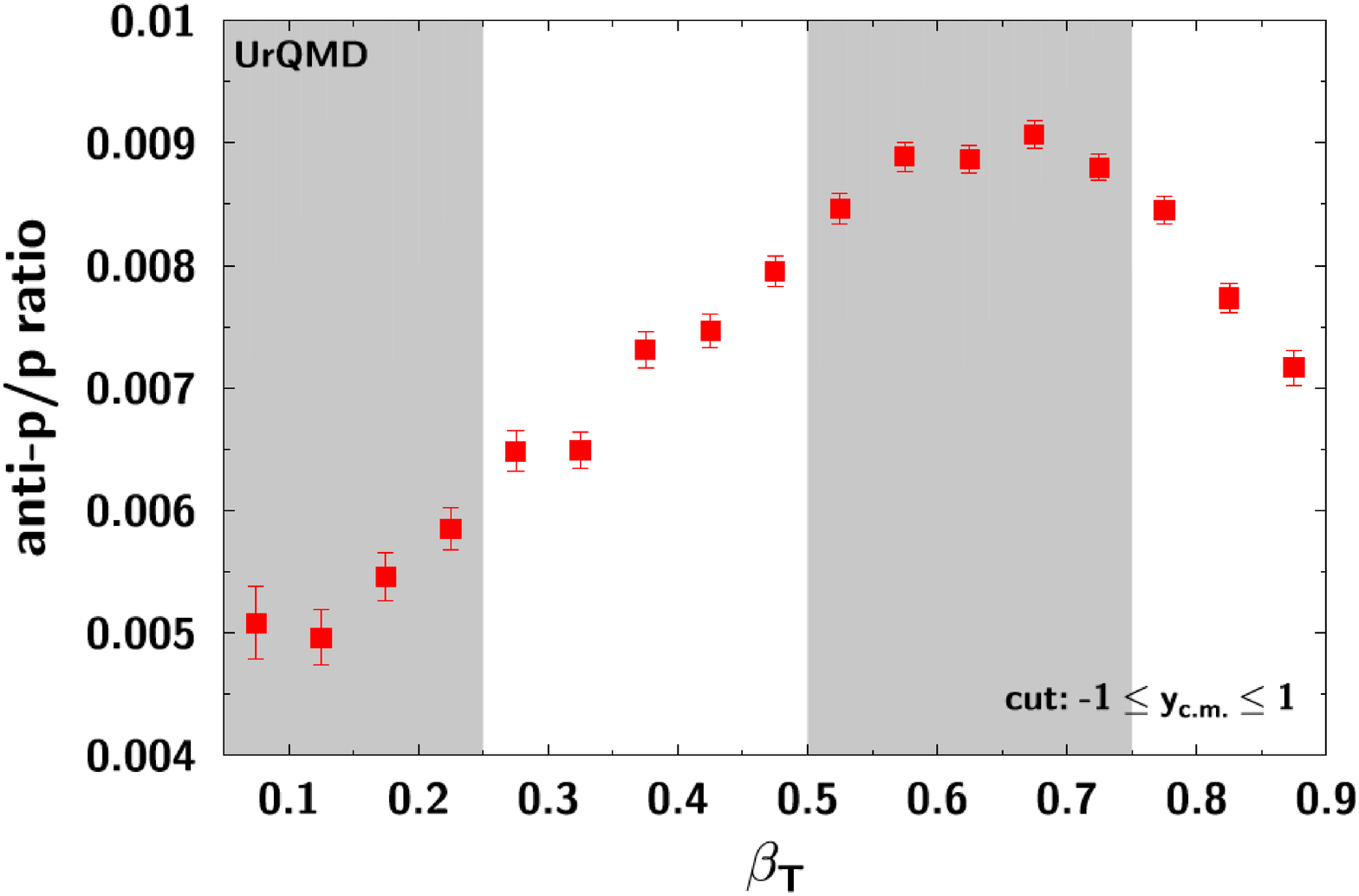}
\caption[]{$\bar{p}/p$ ratio as a function of transverse velocity which is obtained with 
UrQMD. 
}
\label{Fig-ratio-urqmd}
\end{minipage}
\end{figure}

Figure \ref{Fig-xi} shows behavior of static and dynamical fluctuations in 1d hydrodynamic expansion along 
isentropic trajectories on the $\mu_B$-$T$ plane \cite{NoAs05}. 
In both cases of static and dynamic fluctuations, we can see the effect of QCP,  i.e  enhancement of fluctuation 
around QCP.  For static case fluctuation becomes  maximum just at QCP.    However for dynamic case 
maximum value of fluctuation appears after passing in the vicinity of  QCP because of the critical slowing down and 
the maximum value itself  is not so large as the static case. 
There is possibility that fluctuations which are induced by QCP do not become so large as 
a signal of QCP, if the expansion of fireball is fast. 

For hadron ratios key issue from the point of view of QCP is that a chemical freezeout temperature 
depends on transverse velocity of hadrons. 
Figure  \ref{Fig-ratio-urqmd} shows that $\bar{p}/p$ ratio has transverse velocity dependence 
in a microscopic transport model, UrQMD in which the QCP does not exist. 
We find that on  isentropic trajectories the  freezeout process occurs gradually \cite{QCP08}: 
particles with higher transverse velocity  are emitted at earlier time  of expansion and those with lower transverse velocity 
are produced at later time. 
This suggests that hadron ratios  may change on isentropic trajectory between a hadronization point on the QCD phase 
boundary on the $\mu_B$-$T$ plane and chemical freezeout point. 
The hadron ratio,  especially  $\bar{p}/p$ ratio as a function of transverse velocity (momentum) is sensitive to behavior of 
isentropic trajectories on $\mu_B$-$T$ plane and may show a consequence of QCP clearly.  

\begin{figure}[ht]
\centering
\begin{minipage}{0.45\textwidth}
\includegraphics[width=\textwidth]{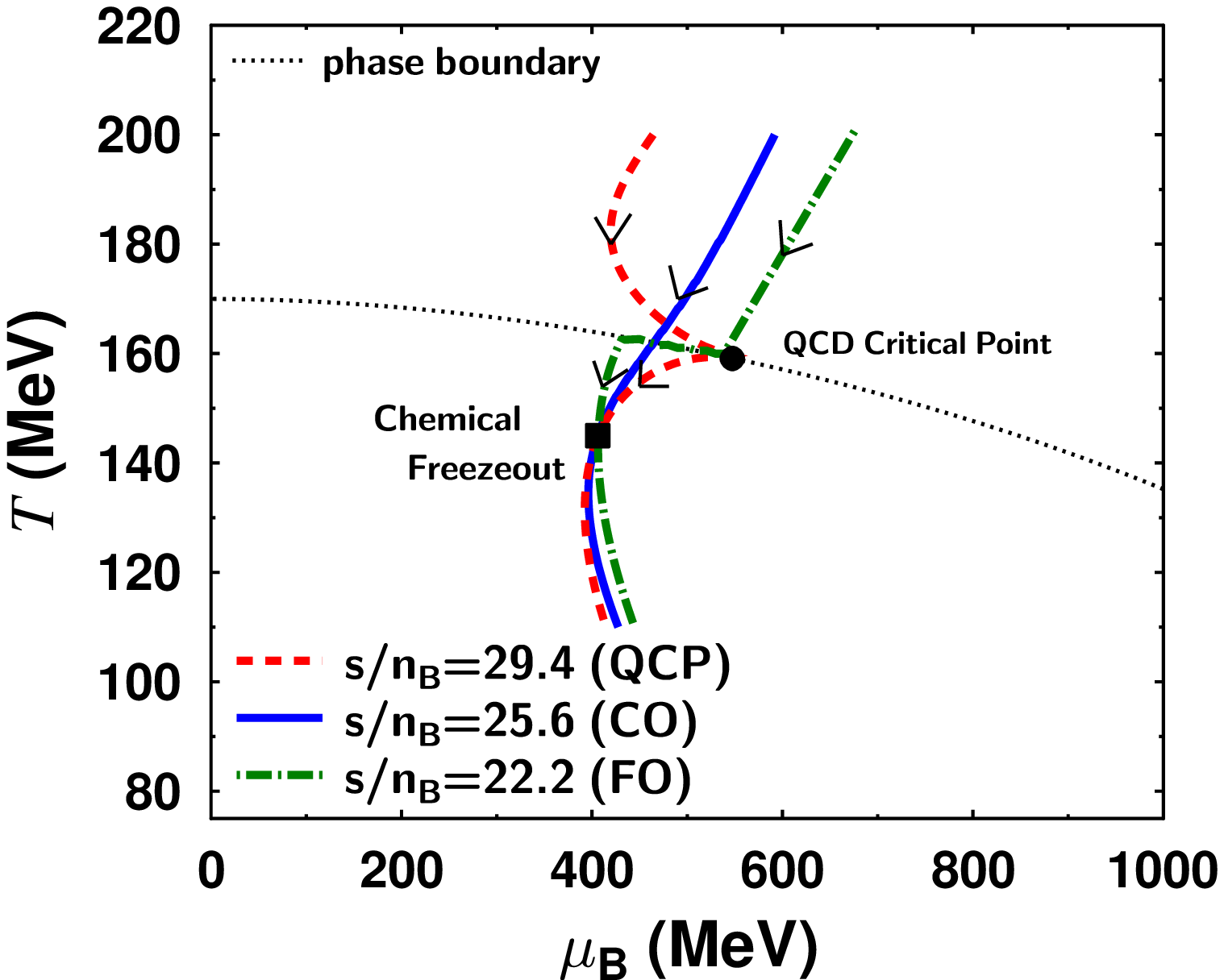}
\caption[]{
Isentropic trajectories with and without QCP on the QCD 
phase diagram. Arrows indicate the direction of time evolution.
}
\label{Fig-tra}
\end{minipage}
\hspace{0.04\textwidth}
\begin{minipage}{0.45\textwidth}
\includegraphics[width=\textwidth]{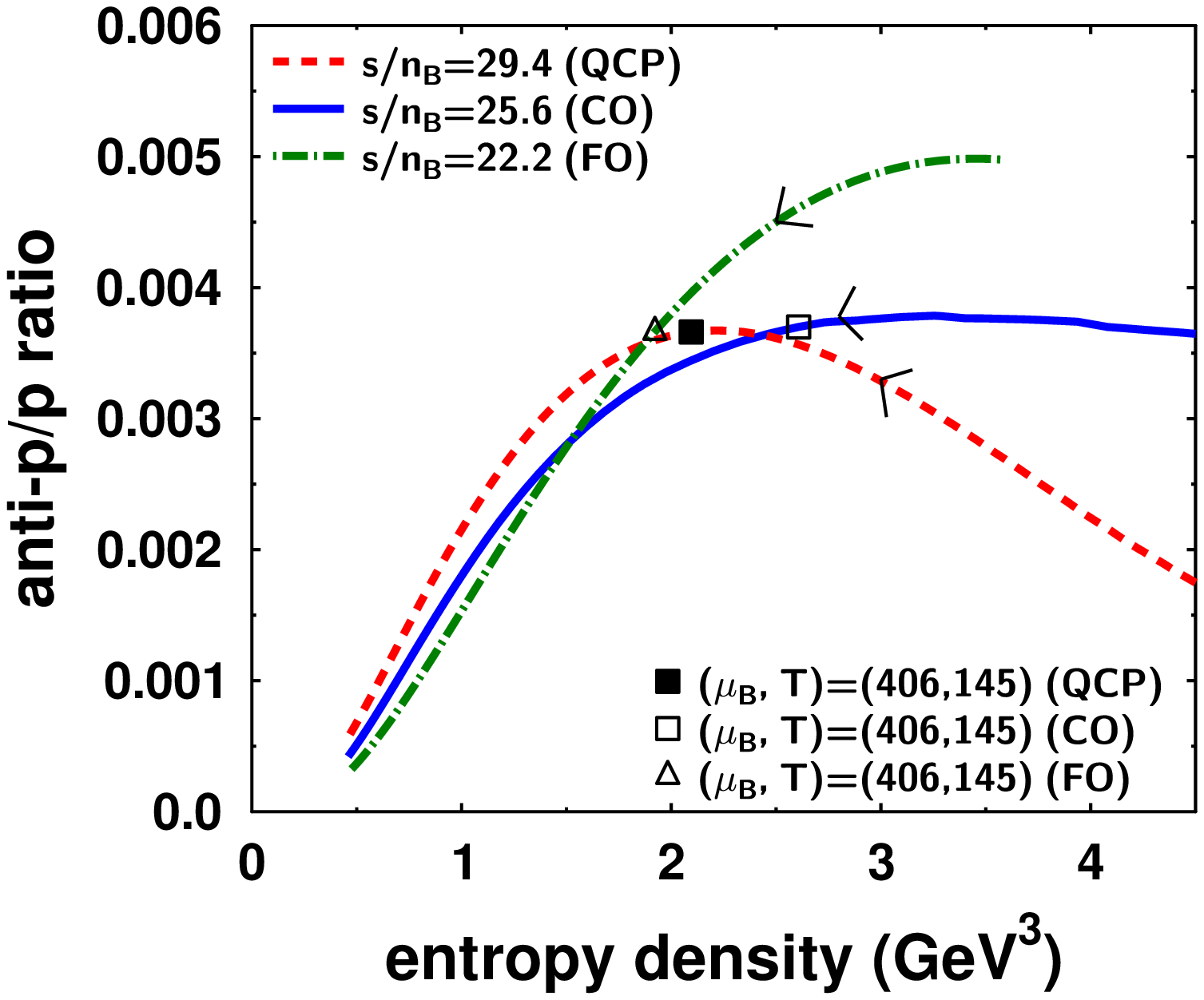}
\caption[]{Antiproton-to-proton ratio along the
trajectories as a function of the entropy density.  Arrows indicate the direction of time evolution.
}
\label{Fig-ratio}
\end{minipage}
\end{figure}

Next we do a demonstrative calculation to show how the signal of QCP appears 
through $\bar{p}/p$ ratio in heavy ion collisions. 
Here we focus on SPS energy region and put the QCP which is parameter in 
our model and the chemical freezeout point which is obtained in a statistical model \cite{sta}  to 
$(\mu_B, T)$ = (550, 159) MeV and (406, 105) MeV on the $\mu_B$-$T$ plane, respectively. 
Figure \ref{Fig-tra} shows that hydrodynamical trajectories in the QCD
phase diagram with and without the presence of a critical point.
Possible trajectories in the  plane in the absence of a
critical point are shown as solid line (for a crossover transition (CO))
or dash-dotted line (for a first-order transition (FO)); the trajectory in
the presence of a critical point is shown as dashed line (QCP). All
trajectories meet at the bulk chemical freezeout point. 
This dotted line in which clear focusing effect appears stands for isentropic trajectory of the QCP. 
Hadronization occurs between the phase boundary and chemical freezeout point. 
And between them we can see clear differences in three cases. 
In the case of QCD critical point the ratio of $\bar{p}{p}$ decreases  along the line or almost the same. 
On the other hand,  in the case of 1st order phase transition and crossover this value 
increases along isentropic trajectory.

Antiproton-to-proton ratio along the
trajectories is shown in Fig. \ref{Fig-ratio} as a function of the entropy density which 
is proportion to transverse momentum.
The curves start at the phase boundary 160 MeV and
continue down to chemical freezeout temperature (145 MeV). 
The location of the chemical freezeout point deduced from experimental data is
indicated by the open and solid squares. Note that the ratio
only rises for the trajectory deformed by the critical point.
In actual experimental data this evidence should appear as steeper $\bar{p}$ spectra at 
high $P_T$.

We find interesting experimental data which may suggest a signal of QCP: 
$\bar{p}$ spectra obtained by NA49 \cite{NA49}. 
They show  $\bar{p}$ and $p$ spectra on collision energies from 20 GeV to 158 GeV. 
Only at 40 GeV collision energy slope of $P_T$ spectra of $\bar{p}$ seems to be steeper 
compared to other collision energies, which would require a trajectory of the type expected 
in the  vicinity of the QCP (Fig. \ref{Fig-ratio}).  The size of the statistical errors of the measurement 
does not permit a firm conclusion about this anomaly, but it is certainly compatible with the 
arguments presented here. 

\begin{figure}[ht]
\centering
\begin{minipage}{0.45\textwidth}
\includegraphics[width=\textwidth]{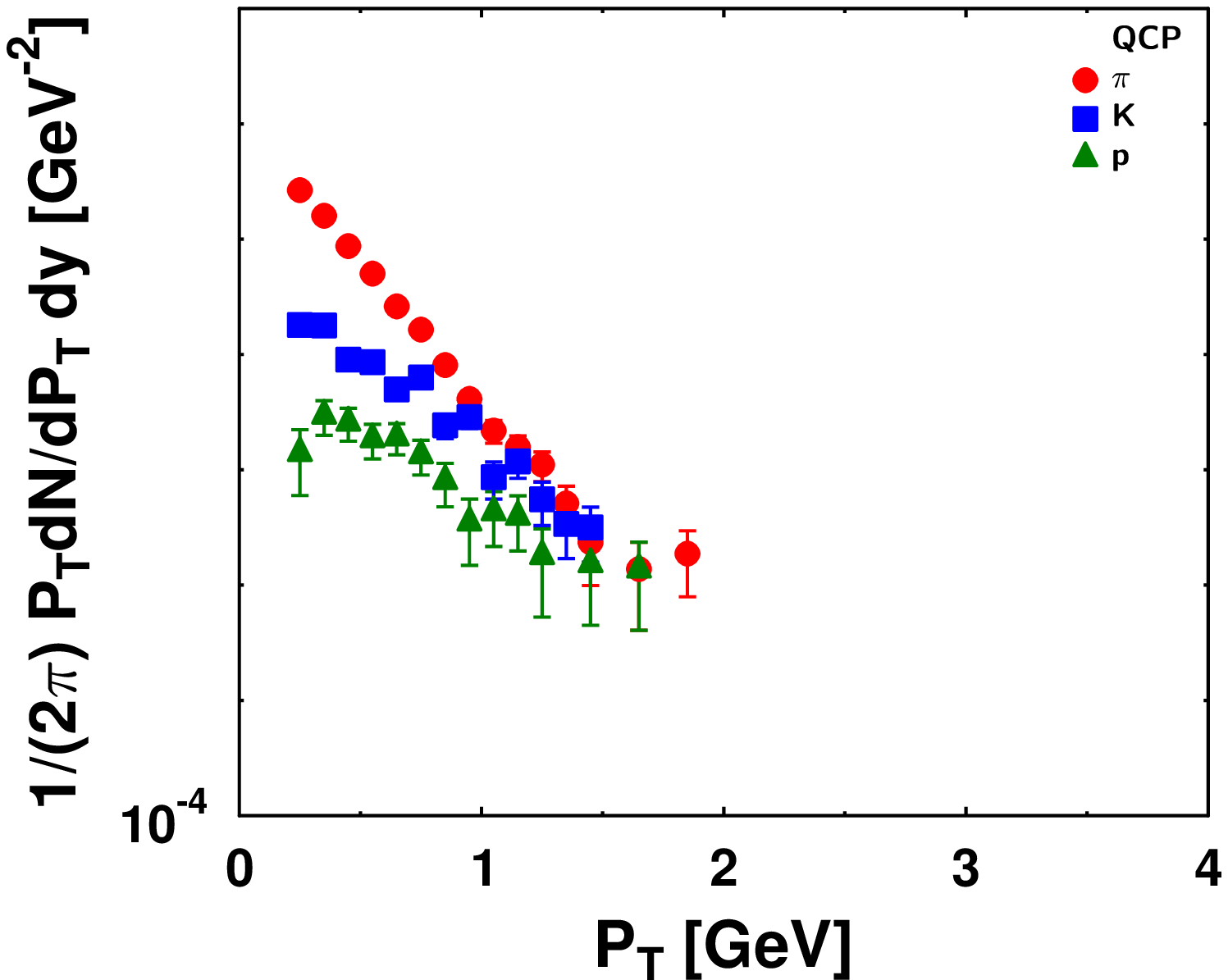}
\end{minipage}
\hspace{0.04\textwidth}
\begin{minipage}{0.45\textwidth}
\includegraphics[width=\textwidth]{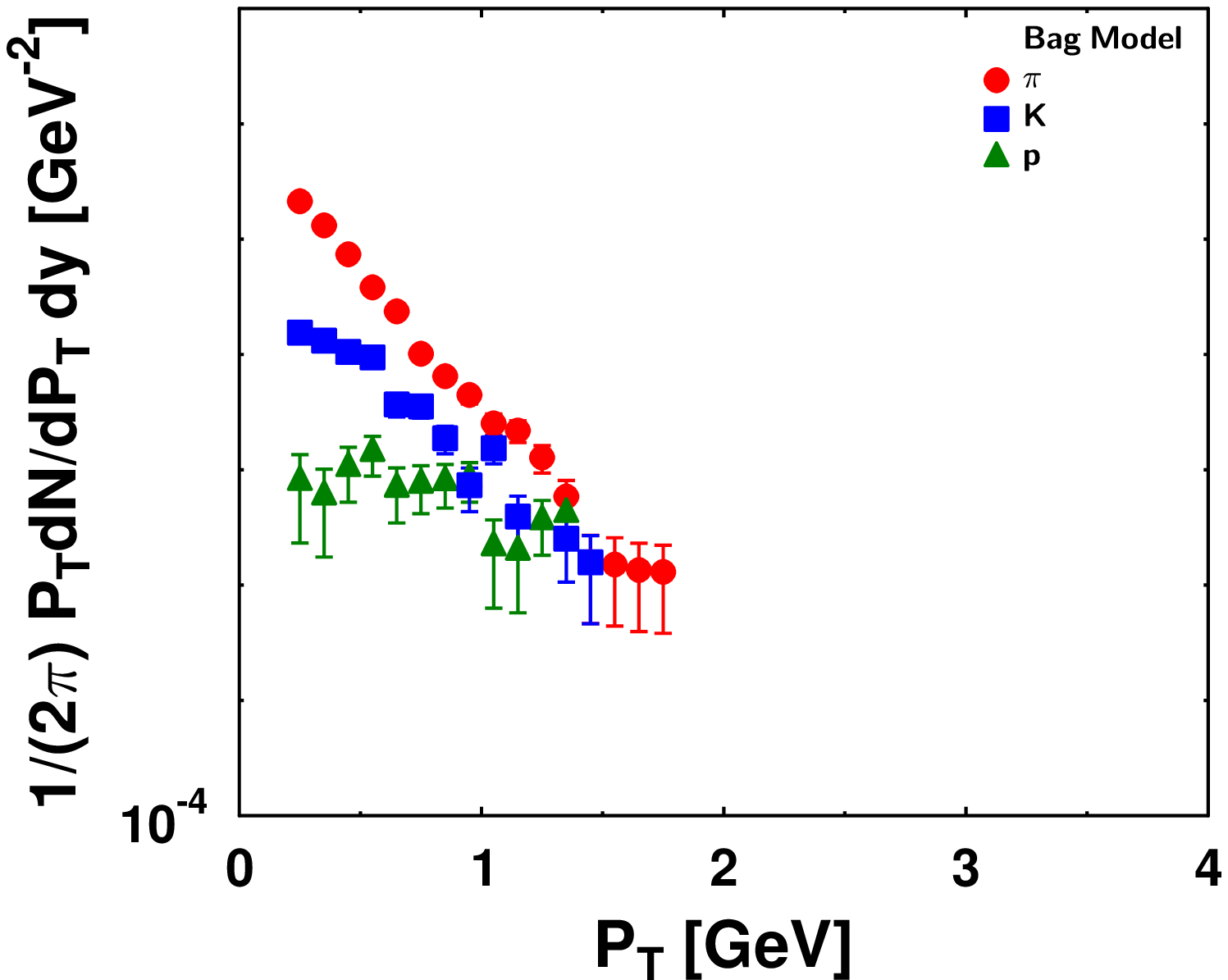}
\end{minipage}
\caption[]{$P_T$ spectra for $\pi$ (solid circles), $K$ (solid square) and 
$p$ (solid triangle) with QCP (left) and without QCP (right). 
}
\label{Fig-pt-spectra}
\end{figure}

Finally we show an example of realistic calculation by 3d hydro + UrQMD model with EoS including QCP. 
For initial conditions of our hybrid model we set maximum value of energy density and baryon 
number density  to be 2.0 GeV/fm$^3$ and 0.15 fm$^{-3}$ , respectively, 
which corresponds to SPS energy region. In the following calculations we use the same 
initial conditions for the both cases of EoS with and without QCP and set a switching temperature 
from hydrodynamic model  to UrQMD to be 150 MeV. 
Because of different behavior in isentropic trajectories between EoS with and without QCP, 
chemical potential at the switching temperature is 430 MeV in presence of QCP
 which is larger than one in absence of QCP (250 MeV). 
This difference appears in hadron ratio. Figure \ref{Fig-pt-spectra} shows $P_T$ spectra of 
$\pi$, $K$, $p$ with QCP (left)  and without QCP (right). 
We can see the effect of different isentrpic trajectoris in hadron ratios. 

In summary, we have shown that the evolution of the 
$\bar{p}/p$ ratio along isentropic curves between the phase 
boundary in the QCD phase diagram and the chemical 
freeze-out point is strongly dependent on the presence or 
absence of a critical point. 
When a critical point exists, the isentropic trajectory 
approximately corresponding to hydrodynamical 
expansion is deformed, and the $\bar{p}/p$ 
ratio grows during the approach to chemical freeze-out.  
Depending on the actual size of the attractive region around the critical point, the search for 
an anomaly in the $P_T$ dependence of the $\bar{p}/p$ 
ratio may require small beam energy steps.

\end{document}